# Balancing the effective sample size in prior across different doses in the curve-free Bayesian decision-theoretic design for dose-finding trials

Jiapeng Xu, MS[1]

Dehua Bi, PhD[1]

Shenghua Kelly Fan, PhD[2]

Bee Leng Lee, PhD[3]

Ying Lu, PhD[1]

1. Department of Biomedical Data Science and Center for Innovative Study Design, School of Medicine, Stanford University, Stanford, CA 94305, USA
2. Department of Statistics and Biostatistics, California State University at East Bay, Hayward, CA 94542, USA
3. Department of Mathematics and Statistics, San José State University, San José, CA 95192, USA

Corresponding author: Ying Lu, PhD, Professor of Biomedical Data Science, Stanford University School of Medicine, Stanford, CA94305-5464, USA. E-mail: ylu1@stanford.edu; Tel: 650-736-8300

**Abstract**

The primary goal of dose allocation in phase I trials is to minimize patient exposure to subtherapeutic or excessively toxic doses, while accurately recommending a phase II dose that is as close as possible to the maximum tolerated dose (MTD). Fan et al. (2012) introduced a curve-free Bayesian decision-theoretic design (CFBD), which leverages the assumption of a monotonic dose-toxicity relationship without directly modeling dose-toxicity curves. This approach has also been extended to drug combinations for determining the MTD (Lee et al., 2017). Although CFBD has demonstrated improved trial efficiency by using fewer patients while maintaining high accuracy in identifying the MTD, it may artificially inflate the effective sample sizes for the updated prior distributions, particularly at the lowest and highest dose levels. This can lead to either overshooting or undershooting the target dose. In this paper, we propose a modification to CFBD's prior distribution updates that balances effective sample sizes across different doses. Simulation results show that with the modified prior specification, CFBD achieves a more focused dose allocation at the MTD and offers more precise dose recommendations with fewer patients on average. It also demonstrates robustness to other well-known dose finding designs in literature.

# Introduction

Phase I oncology dose-finding trials aim to determine the maximum tolerated dose (MTD)—the dose that corresponds to a dose-limiting toxicity (DLT) rate closest to a pre-specified acceptable level. Fan et al. (2012) introduced a curve-free Bayesian decision-theoretic design (CFDB) to determine the MTD for single-agent dose-finding trials. The design only assumes a monotonic dose-toxicity relationship without requiring specification of a dose-toxicity curve. CFBD is iterative and adaptive, beginning with the specification of the prior distributions for DLT rates. As trial data are collected, posterior distributions are calculated and used as updated priors for subsequent decisions, a process we refer to as "prior updates". The MTD is estimated using an isotonic procedure that incorporates both the observed DLT rate at a dose and information extrapolated from other doses. Simulation studies by Fan et al. (2012) demonstrated that CFBD is more likely to identify the target dose with fewer patients compared to other commonly used Bayesian methods, such as the Continual Reassessment Method (CRM). Additionally, CFBD demonstrated robustness against initial prior misspecifications.

Lee et al. (2017) extended CFBD from single-agent trials to two-agent trials by introducing a partial order on the dose combinations. This extension allows for the extrapolation of information among doses while maintaining the partial order constraint for posterior means. Simulation results showed that their design compares favorably in most scenarios to other designs, such as the generalized CRM (gCRM) proposed by Braun and Jia (2013), the YY09a proposed by Yuan and Yin (2009), and the YY09b proposed by Yin and Yuan (2009). Fan et al. (2025) provided a comprehensive review of CFBD, along with a software package for determining the MTD in both single-drug and multi-drug combination trials, and for identifying the optimal effective doses.

The main idea of CFBD is that patients who experience a DLT at a given dose are expected to experience DLTs at higher doses, while those without DLTs at a given dose are unlikely to experience DLTs at lower doses. By extrapolating these unobserved outcomes, information can be borrowed from the current dose level to update the priors for the lower and the higher doses not tested in the current cohort. This process effectively increases the information available for all dose levels beyond the currently tested level. While this extrapolation allows for quick updates to the estimates of toxicity rates across different dose levels, it risks over-representing the available information, especially at the highest and lowest doses. This may result in the prior at the lowest and highest doses having a larger effective sample size, inadvertently inflating the influence of the extrapolated information. Although the estimate for the DLT rate remains unchanged, an inflated effective sample size can alter the shape of the prior distribution, which in turn affects the utility of each dose level. Consequently, patients may be allocated to, or recommended for, doses that are either subtherapeutic or excessively toxic, thereby exacerbating the over/undershooting issue. Addressing this over/undershooting issue is crucial to the integrity and success of phase I trials. Inaccurate dose allocation can result in failure to identify the optimal therapeutic dose, leading to either ineffective treatment or unnecessary toxicity for patients.

In this paper, we propose a modification of original CFBD, termed c-CFBD, which balances the effective sample sizes in prior updates via calibration. Our method is straightforward to implement and ensures more conservative borrowing across doses by maintaining a consistent monotonic prior mean DLT rates without inflating the effective sample sizes in the updated priors. Simulation results demonstrate that, in most scenario, this calibration method leads to more concentrated dose allocations around the MTD and more accurate dose recommendations, while requiring fewer participants on average. These improvements enhance the efficiency and accuracy of phase I trials, ensuring better resource utilization.

In addition to the CFBD, there are other Bayesian phase I designs for the identification of MTD. The most popular among these are the Bayesian optimal interval (BOIN) design (Liu et al., 2015; Lin et al., 2017) and the modified toxicity probability interval (mTPI2) design (Yan et al., 2017). These methods have substantially improved the performance of phase I trials and are now considered the standard practice. Therefore, in this paper, we also compare c-CFBD with both BOIN and mTPI2 designs.

The rest of this paper is organized as follows. Section 2 briefly introduces the original CFBD and proposes the proposed c-CFBD design to balance effective sample sizes in prior updates. Section 3 reports the simulation results comparing the proposed method with the original design. Finally, Section 4 concludes with a discussion of the findings.

# 1 Balancing the effective sample size

## 1.1 Review of CFBD for one- and two-agents

The one- and two-agent CFBD designs use an isotonic working model with conjugate beta priors to reduce computational burden. Consider a single agent trial with binary DLT response. Assume there are $J$ dose levels, where $n_j$ denotes the number of patients treated at dose $j$, and $t_j$ is the number of DLTs observed at this dose level. Under the conjugate Beta-binomial model, we have $t_j|n_j, p_j \sim \text{Binom}(p_j, n_j)$, $p_j \sim \text{Beta}(a_j, b_j)$ for $j = 1, ..., J$, where $p_j$ is the DLT probability in dose $j$, and $a_j, b_j$ are the hyperparameters of the Beta prior distribution. The CFBD design updates $a_j$ and $b_j$ after each cohort of patients. Specifically, the initial Beta parameters for each dose levels prior to the start of the trial, denoted as $a_j^{(0)}$ and $b_j^{(0)}$, must be chosen to ensure monotonicity of the mean DLT rates across doses. Once the trial begins and data is collected, the design extrapolates working data to update the hyperparameters of the Beta prior while maintaining the monotonic property in DLT rates. The process is described as follows. Let $a_j^{(k-1)}$ and $b_j^{(k-1)}$ denote the posterior hyperparameters for dose $j$, calculated from all data collected after the $(k-1)$-th cohort of patients. These hyperparameters also serve as the prior hyperparameters for the $k$th cohort. Moreover, assume the previous decision is made to proceed to dose $j'$, where $n_{j'}$ patients are enrolled and treated at this dose level and $t_{j'}$ DLTs are observed. Following the method of Fan et al. (2012), the hyperparameters for all dose levels are updated after the data from the $k$th cohort are collected using the following equations:

$$\text{Beta}\left(a_j^{(k)}, b_j^{(k)}\right) = \begin{cases} \text{Beta}\left(a_j^{(k-1)}, b_j^{(k-1)} + n_{j'} - t_{j'}\right) & \text{if } j < j' \\ \text{Beta}\left(a_j^{(k-1)} + t_{j'}, b_j^{(k-1)} + n_{j'} - t_{j'}\right) & \text{if } j = j' \\ \text{Beta}\left(a_j^{(k-1)} + t_{j'}, b_j^{(k-1)}\right) & \text{if } j > j' \end{cases} \quad (1)$$

for $j = 1, \ldots, J$. This method of data extrapolation maintains the monotonicity of the posterior means with respect to the dose levels.

The dose assignment rule of CFBD selects the dose level for the next cohort of patients using the one-step-look-ahead (OSLA) method to optimize a utility function. Specifically, the utility function is composed of an individual gain function and a weight distribution over all applicable individuals. The individual gain function is selected based on the primary goal of the trial, while the weight distribution is chosen to reflect how benefits are balanced between in-trial patients and future patients. In this paper, our primary aim is to identify the MTD as frequently as possible while prioritizing the gains of future patients over those currently in the trial. A suitable individual gain function might take the form of $-\left|p_j^{(k)} - \theta_0\right|$, where $p_j^{(k)}$ is the DLT rate of assigning the $k$th cohort to dose $j$ and $\theta_0$ is the target DLT rate, and the corresponding weight distribution assigns a weight of 1 to future patients and 0 to in-trial patients. Once the individual gain function and weight distribution are specified, the utility function is given by the total weighted gain from all in-trial and future patients. Since finding the optimal dose is computationally infeasible in most situations and may not always be better because of some practical requirements, Fan et al. (2012) uses a commonly used suboptimal rule known as the one-step-look-forward (OSLA), which assumes that the current patient is the last one available and selects the best-so-far dose. In this approach, let $u_k\left(p_j^{(k)}, \theta_0\right)$ denote the utility for assigning the $k$th cohort to dose $j$ and let $\text{Data}_k$ denote the data collected before the $k$th cohort. The expected utility function is simply $E\left(u_k\left(p_j^{(k)}, \theta_0\right) \big| \text{Data}_k\right)$, with the optimal dose being the one that maximizes this expectation, i.e., $\text{argmax}_j\, E\left(u_k\left(p_j^{(k)}, \theta_0\right) \big| \text{Data}_k\right)$.

Lastly, four stopping rules are used in the CFBD design in one-agent:

1) Early stopping before reaching the minimum sample size $n_{min}$ is prohibited.
2) The maximum sample size $n_{max}$ is reached.
3) All doses are deemed too toxic, that is, $P[p_1 > \theta_0 + \delta_0 | data] > r_1$, where $\delta_0$ is a prespecified constant such that $\theta_0 + \delta_0$ *is an upper bound on the acceptable DLT* rate, and $r_1$ is a prespecified probability.
4) The current MTD is very likely to be overly toxic, i.e., if $P[p_{\text{MTD}^+} > \theta_0 + \delta_0 | data] > r_2$, where $\text{MTD}^+$ is the next higher dose from the current estimated MTD.

Among the four stopping rules, rules 1 and 2 take precedence over rules 3 and 4.

For two-agent trials, Lee et al. (2017) extends the CFBD design by introducing a partial order to establish the monotonicity relationship between dose combinations. Assume there are $I$ and $J$ dose levels in agents 1 and 2, respectively. Let $p_{ij}$ denote the DLT rate at dose combination $(i, j)$, where $i \in \{1, \ldots, I\}$ and $j \in \{1, \ldots, J\}$. Combination $(i, j)$ is considered less toxic than combination $(i', j')$, denoted $(i, j) \prec (i'j')$, if and only if $p_{ij} \leq p_{i'j'}$, or equivalently, when $i \leq i'$, $j \leq j'$, and $i + j \leq i' + j'$; otherwise, $(i, j)$ and $(i', j')$ are not comparable. Based on this partial ordering, the CFBD design for two agents uses the same Beta-binomial model but updates the hyperparameters differently. Let $a_{ij}^{(k-1)}$ and $b_{ij}^{(k-1)}$ denote the hyperparameters for the Beta prior for the $k$th cohort, obtained from the data up to the $(k-1)$th cohort. Further assume the patients are currently treated at dose level $(r, s)$, where $n_{rs}$ patients are enrolled and $t_{rs}$ DLTs are observed. Similar to the method of Fan et al. (2012), Lee et al. (2017) updates $a_{ij}^{(k)}$ and $b_{ij}^{(k)}$ for all $i = 1, \ldots, I$ and $j = 1, \ldots, J$ as follows:

$$Beta\left(a_{ij}^{(k)}, b_{ij}^{(k)}\right) = \begin{cases} Beta\left(a_{ij}^{(k-1)}, b_{ij}^{(k-1)} + n_{rs} - t_{rs}\right) & if\ (i,j) \prec (r,s) \\ Beta\left(a_{ij}^{(k-1)} + t_{rs}, b_{ij}^{(k-1)} + n_{rs} - t_{rs}\right) & if\ i = r\ and\ j = s \\ Beta\left(a_{ij}^{(k-1)} + t_{rs}, b_{ij}^{(k-1)}\right) & if\ (i,j) \succ (r,s) \\ Beta\left(a_{ij}^{(k-1)}, b_{ij}^{(k-1)}\right) & Otherwise \end{cases} \quad (2)$$

Lee et al. (2017) adopted the same utility function and dose selection rule to two-agent trials. They introduced two tuning parameters, $\alpha_0 > 0$ and $\eta_0 > 0$, which control the utility for under- and over-dosing, respectively. Let $p_{ij}^{(k)}$ denote the DLT rate when assigning patient $k$ to dose $(i,j)$, and $\theta_0$ be the target DLT rate. Lee et al. (2017) defined the utility function $u_k\left(p_{ij}^{(k)}, \theta_0\right)$ as:

$$u_k\left(p_{ij}^{(k)}, \theta_0\right) = \begin{cases} -\alpha_0\left(\theta_0 - p_{ij}^{(k)}\right); & if\ p_{ij}^{(k)} \leq \theta_0 \\ -\eta_0\left(p_{ij}^{(k)} - \theta_0\right); & if\ p_{ij}^{(k)} > \theta_0 \end{cases}$$

Similar to Fan et al. (2012), the next dose assignment is determined by the dose that results in the highest expected utility, $\text{argmax}_{(i,j)} E\left(u_k\left(p_{ij}^{(k)}, \theta_0\right) \mid \text{Data}_k\right)$. Let $f\left(p; a_{ij}^{(k)}, b_{ij}^{(k)}\right)$ and $F\left(p; a_{ij}^{(k)}, b_{ij}^{(k)}\right)$ denote the density and cumulative distribution function of a $\text{Beta}\left(a_{ij}^{(k)}, b_{ij}^{(k)}\right)$, where $a_{ij}^{(k)}$ and $b_{ij}^{(k)}$ are the parameters of the prior distribution for $p_{ij}^{(k)}$ given all data collected up to the $k$th cohort. Then

$$E\left[u_k\left(p_{ij}^{(k)}, \theta_0\right) \mid a_{ij}^{(k)}, b_{ij}^{(k)}\right]$$

$$= \int_0^{\theta_0} -\alpha_0\left(\theta_0 - p_{ij}^{(k)}\right) f\left(p_{ij}^{(k)}; a_{ij}^{(k)}, b_{ij}^{(k)}\right) dp_{ij}^{(k)}$$

$$+ \int_{\theta_0}^1 -\eta_0\left(p_{ij}^{(k)} - \theta_0\right) f\left(p_{ij}^{(k)}; a_{ij}^{(k)}, b_{ij}^{(k)}\right) dp_{ij}^{(k)}$$

$$= -(\alpha_0 + \eta_0)\left[\theta_0 F\left(\theta_0; a_{ij}^{(k)}, b_{ij}^{(k)}\right) - \frac{a_{ij}^{(k)}}{\left(a_{ij}^{(k)} + b_{ij}^{(k)}\right)} F\left(\theta_0; a_{ij}^{(k)} + 1, b_{ij}^{(k)}\right)\right] - \eta_0\left[\frac{a_{ij}^{(k)}}{\left(a_{ij}^{(k)} + b_{ij}^{(k)}\right)} - \theta_0\right]$$

The next dose assignment is given by the dose combination $(i, j)$ that maximizes $E\left(u_k\left(p_{ij}^{(k)}, \theta_0\right) \mid Data_k\right)$.

Lastly, for the stopping rules, while rules 1 and 2 remain the same, rules 3 and 4 are modified. The rule 3 becomes $P[p_{11} > \theta_0 + \delta_0 | data] > r_1$ (i.e., all doses are deemed too toxic) and the rule 4 becomes $\min_{(i,j) \prec (r,s)} P(p_{rs} > \theta_0 + \delta_0 | data) > r_2$ (i.e., the current MTD is very likely to be overly toxic).

## 1.2 Proposed calibration

Although the isotonic working model can enrich the information obtained from an individual response, after each update, the effective sample size in each dose level would increase differently, resulting in a potential to over-represent the available information at certain doses. To address this issue, we propose to calibrate the hyperparameters to maintain a same effective sample size across the dose levels, with the effective sample size to be maintained being the average across all dose levels.

Specifically, consider a one-agent dose finding trial with $J$ doses in total. After assigning the $(k-1)$-th cohort of patients to dose level $j'$, we propose to calibrate the hyperparameters $a_j^{(k)}$ and $b_j^{(k)}$ in CFBD by $l_j^{(k)}$, which is computed as a ratio of dose $j$'s effective sample size to the average current effective sample size across all doses, denoted as $S^{(k)}$. Mathematically, we have

$$S^{(k)} = \frac{\sum_{j=1}^{J} \left(a_j^{(k)} + b_j^{(k)}\right)}{J} \quad \text{and} \quad l_j^{(k)} = \frac{a_j^{(k)} + b_j^{(k)}}{S^{(k)}},$$

where $a_j^{(k)}$ and $b_j^{(k)}$ are specified in Equation (1), and the calibrated hyperparameters, denoted as $\alpha_j^{(k)}$ and $\beta_j^{(k)}$ are given by the follows:

$$\alpha_j^{(k)} = \frac{a_j^{(k)}}{l_j^{(k)}} \quad \text{and} \quad \beta_j^{(k)} = \frac{b_j^{(k)}}{l_j^{(k)}}, for\ j = 1, \dots, J.$$

In other words, the $Beta\left(\alpha_j^{(k)}, \beta_j^{(k)}\right)$s are the updated priors for the subsequent dose cohort in the calibrated design, which we called c-CFBD. Under this calibration, the dose levels with an effective sample size that is greater than the average will be down-weighted while those that is less than the average will be up-weighted to maintain a same effective sample size across all dose levels, while maintaining the mean and monotonicity of the hyperparameters in the CFBD design. After this calibration, the c-CFBD design uses the same dose assignment rules, early stopping rules, and method to find MTD as the CFBD design.

For two-agents CFBD, we use similar calibration idea to calibrate the Beta hyperparameters for $a_{ij}^{(k)}$ and $b_{ij}^{(k)}$ in equation (2). Assume there are $I$ and $J$ dose levels in agents 1 and 2, respectively, and the $k$th cohort of patients are assigned to dose level $(r,s)$. We can compute $a_{ij}^{(k)}$ and $b_{ij}^{(k)}$ for all dose levels $i$ and $j$ as in equation (2). The average effective sample size, $S^{(k)}$, and the ratio $l_{ij}^{(k)}$ can then be computed as follows:

$$S^{(k)} = \frac{\sum_{i=1}^{m_1}\sum_{j=1}^{m_2}\left(a_{ij}^{(k)} + b_{ij}^{(k)}\right)}{I*J}, \qquad l_{ij}^{(k)} = \frac{a_{ij}^{(k)} + b_{ij}^{(k)}}{S^{(k)}}$$

Lastly, the calibrated hyperparameters, $\alpha_{ij}^{(k)}$ and $\beta_{ij}^{(k)}$, are given as follows:

$$\alpha_{ij}^{(k)} = \frac{a_{ij}^{(k)}}{l_{ij}^{(k)}} \text{ and } \beta_{ij}^{(k)} = \frac{b_{ij}^{(k)}}{l_{ij}^{(k)}}, \qquad for\ i = 1, \ldots, I;\ j = 1, \ldots, J.$$

It is straightforward to see that the effective sample size after the calibration has also been kept constant across each dose levels. In other words, the calibration process adjusts the influence of extrapolated data to prevent over-representation while preserving estimates of the DLT rate as same. This reduces the utility of doses that deviate from the true MTD and helping to avoid assigning patients to subtherapeutic or excessively toxic doses. The rest of the design, including dose assignment rules, early stopping rules, and method to find the MTD follows the CFBD design of Lee et al. (2017).

# 2 Comparison via simulations

In this section, we compare how the proposed calibration impacts the operating characteristics of the designs proposed by Fan et al. (2012) and Lee et al. (2017). Additionally, we also evaluate the operating characteristics of the proposed design by comparing it with established Phase I clinical trial designs, including BOIN for single-agent trials (Liu and Yuan, 2015) and two-agent trials (Lin and Yin, 2017), as well as mTPI-2/Keyboard for both single- and two-agent trials (Yan et al., 2017). To ensure fair comparisons, we also examine designs using additionally early stopping rules when adequate certainty is available for the selected MTD, which was not available in standard BOIN or mTPI2 designs. The simulations were conducted using open-source packages available from the R CRAN repository. The trial goal is set to identify the MTD as frequently as possible while maintaining a more concentrated dose allocation at the true MTD and minimizing the average sample size. Additionally, the four stopping rules of CFBD and CFBD-2 are used, with rules 1) and 2) having a higher priority than 3) and 4).

For BOIN and mTPI2/Keyboard designs, stopping rules 2) to 3) are used.

## 2.1 Simulation study for one-agent trials

The six scenarios for the single-agent trial (Table 1) were selected to be the same as those in Fan et al. (2012), with a minor modification in the medium scenario. Each simulation was repeated 5,000 times. The minimum sample size $n_{min}$ is 10 and the maximum sample size $n_{max}$ is 24. In the first three scenarios, the target level ($\theta_0$) is chosen to be 0.30, and the maximum tolerated toxicity ($\theta_0 + \delta_0$) is chosen to be 0.35. For the last three scenarios, the target level and maximum tolerated toxicity are 0.20 and 0.25, respectively.

Table 1: True toxicity rates for Scenario 1 to 6, where Scenario 1 refers low toxicity under a 30% target, and each row represents a different scenario. MTD is shown in bold for each scenario.

| Scenario | \multicolumn{6}{c}{dose level} | | | | | |
|---|---|---|---|---|---|---|
|  | 1 | 2 | 3 | 4 | 5 | 6 |
| Target toxicity level: 0.30 | | | | | | |
| Low toxicity | 0.02 | 0.03 | 0.06 | 0.10 | 0.18 | **0.30** |
| Medium toxicity | 0.05 | 0.10 | 0.20 | **0.30** | 0.50 | 0.50 |
| High toxicity | **0.30** | 0.53 | 0.77 | 0.87 | 0.95 | 0.98 |
| Target toxicity level: 0.20 | | | | | | |
| Low toxicity | 0.00 | 0.00 | 0.00 | 0.01 | 0.07 | **0.20** |
| Medium toxicity | 0.01 | 0.02 | 0.09 | **0.20** | 0.50 | 0.68 |
| High toxicity | 0.15 | **0.20** | 0.38 | 0.52 | 0.70 | 0.80 |

The operating characteristics, including dose allocation, dose recommendation, and average sample size, are summarized in Table 2. Comparing to the CFBD design, the most notable improvement is observed in low toxicity scenarios, where the c-CFBD is able to mitigate undershooting and ensure that patients receive therapeutically effective doses. c-CFBD increases the precision of dose recommendation by 3.9% in scenario 1 and by 12% in scenario 4, along with a more concentrated dose allocation towards the true MTD. For medium toxicity scenarios, there is also an increase in precision of dose recommendation (increases by 7.6% in scenario 2 and by 9.5% in scenario 5), accompanied by a decrease in the average sample size. However, in the case of high toxicity scenarios, there is a slight drop in precision and a more even distribution of dose allocation, suggesting that calibration may not be as beneficial in high toxicity environments, where the distribution of the effective sample size in each simulated trial is more likely to align with the true patient allocation that is more concentrated at lower doses.

Comparing to the BOIN and mTPI2/Keyboard designs without the early stopping rule (rule 4) added, the proposed c-CFBD outperforms the other designs both in dose allocation and recommendation in scenarios with high toxicity (Scenarios 3 and 6). It also improves dose allocation in low-toxicity scenarios (Scenarios 1 and 4) and dose recommendation in medium-toxicity scenarios (Scenarios 2 and 5). Moreover, without early stopping rule added, it is clear both the BOIN and the mTPI2/Keyboard designs have higher expected sample sizes than both the CFBD and c-CFBD. As a sensitivity analysis, we added the stopping rule to BOIN and mTIP2/Keyboard, which allows these designs to stop. It is clear from Table 2 that when the rule is added, BOIN and mTPI2/Keyboard designs are able to achieve a lower sample size, which are in general only slightly larger than the expected sample size of the proposed c-CFBD design. However, the c-CFBD design still outperforms both designs with higher probabilities of correctly allocating patients to the right dose and recommending the appropriate dose for later phases.

*Table 2: Results comparing operating characteristics of the proposed c-CFBD, the CFBD design of Fan et al. (2012), BOIN without and with stopping rule 4 (labeled as default BOIN and BOIN+SR, respectively), and mTPI2/Keyboard without and with stopping rule 4 (labeled as Default mTPI2 and mTPI2+SR). Comparisons are shown in terms of dose allocation, dose recommendation, and average sample size (column labeled by E(n)). Operating characteristics under the MTD are shown in bold.*

|  | Dose Allocation | | | | | | Dose recommendation | | | | | | E(n) |
|---|---|---|---|---|---|---|---|---|---|---|---|---|---|
|  | 1 | 2 | 3 | 4 | 5 | 6 | 1 | 2 | 3 | 4 | 5 | 6 |  |
| Scenario 1: Low Toxicity | | | | | | | | | | | | | |
| c-CFBD | 5.5 | 5.4 | 5.5 | 10.2 | 22.9 | **50.5** | 0.0 | 0.0 | 0.2 | 3.4 | 36.6 | **59.8** | 18.2 |
| CFBD | 5.4 | 5.3 | 5.5 | 14.4 | 22.8 | **46.6** | 0.0 | 0.0 | 0.1 | 7.1 | 36.9 | **55.9** | 18.5 |
| Default BOIN | 5.1 | 6.0 | 8.5 | 15.0 | 27.3 | **38.2** | 0.0 | 0.0 | 1.1 | 7.1 | 31.5 | **60.3** | 24.0 |
| BOIN+SR | 5.6 | 6.6 | 9.1 | 14.4 | 23.4 | **41.0** | 0.0 | 0.3 | 1.8 | 9.4 | 35.3 | **53.3** | 21.3 |
| Default mTPI2 | 4.9 | 5.5 | 7.2 | 12.3 | 25.6 | **44.6** | 0.0 | 0.0 | 0.4 | 5.3 | 30.3 | **64.0** | 24.0 |
| mTPI2+SR | 5.4 | 6.1 | 7.7 | 12.1 | 21.5 | **47.2** | 0.0 | 0.1 | 1.2 | 7.7 | 34.3 | **56.7** | 21.3 |
| Scenario 2: Medium Toxicity | | | | | | | | | | | | | |
| c-CFBD | 8.8 | 15.4 | 34.4 | **28.0** | 11.3 | 2.1 | 0.0 | 3.2 | 30.1 | **54.2** | 12.4 | 0.1 | 11.6 |
| CFBD | 7.7 | 18.7 | 35.3 | **28.1** | 8.4 | 1.8 | 0.0 | 6.5 | 37.6 | **46.6** | 9.2 | 0.0 | 13.3 |
| Default BOIN | 8.5 | 16.6 | 26.9 | **28.7** | 13.9 | 5.3 | 0.3 | 6.4 | 30.9 | **45.7** | 14.1 | 2.7 | 24.0 |
| BOIN+SR | 10.3 | 18.0 | 26.0 | **24.4** | 14.2 | 7.1 | 0.9 | 9.8 | 28.9 | **42.4** | 15.1 | 2.9 | 18.6 |
| Default mTPI2 | 6.9 | 13.8 | 25.9 | **31.2** | 15.3 | 6.9 | 0.1 | 5.2 | 28.8 | **47.4** | 14.6 | 3.9 | 24.0 |
| mTPI2 | 9.1 | 15.3 | 24.4 | **26.1** | 16.3 | 8.8 | 0.9 | 7.6 | 27.4 | **44.0** | 16.5 | 3.5 | 17.9 |
| Scenario 3: High Toxicity | | | | | | | | | | | | | |
| c-CFBD | **83.6** | 12.4 | 3.2 | 0.7 | 0.1 | 0.0 | **89.6** | 8.3 | 0.0 | 0.0 | 0.0 | 0.0 | 10.2 |
| CFBD | **86.1** | 9.9 | 3.2 | 0.7 | 0.1 | 0.0 | **92.7** | 5.5 | 0.0 | 0.0 | 0.0 | 0.0 | 10.4 |
| Default BOIN | **73.0** | 22.2 | 4.1 | 0.6 | 0.1 | 0.0 | **73.9** | 11.4 | 0.2 | 0.0 | 0.0 | 0.0 | 22.6 |
| BOIN+SR | **70.5** | 23.1 | 5.5 | 0.9 | 0.1 | 0.0 | **71.8** | 15.0 | 0.7 | 0.1 | 0.0 | 0.0 | 16.6 |
| Default mTPI2 | **70.2** | 24.6 | 4.6 | 0.6 | 0.1 | 0.0 | **73.4** | 12.4 | 0.2 | 0.0 | 0.0 | 0.0 | 22.7 |
| mTPI2+SR | **67.6** | 25.3 | 6.1 | 0.8 | 0.1 | 0.0 | **71.7** | 16.1 | 0.6 | 0.0 | 0.0 | 0.0 | 16.5 |
| Scenario 4: Low Toxicity | | | | | | | | | | | | | |
| c-CFBD | 5.1 | 5.1 | 5.1 | 5.3 | 18.3 | **61.1** | 0.0 | 0.0 | 0.0 | 0.0 | 22.7 | **77.3** | 19.7 |
| CFBD | 5.1 | 5.1 | 5.1 | 6.4 | 29.3 | **49.1** | 0.0 | 0.0 | 0.0 | 0.9 | 33.8 | **65.3** | 19.8 |
| Default BOIN | 4.2 | 4.2 | 4.4 | 7.9 | 26.8 | **52.6** | 0.0 | 0.0 | 0.0 | 0.9 | 26.9 | **72.2** | 24.0 |
| BOIN+SR | 4.8 | 4.8 | 5.1 | 8.0 | 20.0 | **57.2** | 0.0 | 0.0 | 0.0 | 3.2 | 33.5 | **63.3** | 20.7 |
| Default mTPI2 | 4.2 | 4.2 | 4.4 | 7.7 | 25.7 | **54.0** | 0.0 | 0.0 | 0.0 | 0.9 | 25.7 | **73.3** | 24.0 |
| mTPI2+SR | 4.9 | 4.9 | 5.1 | 8.0 | 19.9 | **57.3** | 0.0 | 0.0 | 0.0 | 3.0 | 34.5 | **62.4** | 20.6 |
| Scenario 5: Medium Toxicity | | | | | | | | | | | | | |

| Design | | | | | | | | | | | | |
|---|---|---|---|---|---|---|---|---|---|---|---|---|
| c-CFBD | 8.2 | 10.9 | 33.6 | **38.8** | 5.6 | 2.9 | 0.0 | 0.8 | 21.3 | **76.0** | 1.8 | 0.0 | 12.5 |
| CFBD | 7.2 | 11.0 | 36.5 | **38.4** | 4.6 | 2.3 | 0.1 | 2.8 | 30.6 | **66.5** | 0.0 | 0.0 | 15.4 |
| Default BOIN | 5.0 | 9.6 | 25.0 | **40.1** | 16.3 | 4.0 | 0.0 | 2.2 | 26.3 | **65.1** | 6.4 | 0.1 | 24.0 |
| BOIN+SR | 7.8 | 12.0 | 24.1 | **30.1** | 19.7 | 6.3 | 0.3 | 4.2 | 26.2 | **56.0** | 13.1 | 0.2 | 14.8 |
| Default mTPI2 | 4.9 | 9.5 | 24.7 | **40.6** | 16.3 | 4.0 | 0.0 | 2.1 | 25.7 | **65.7** | 6.2 | 0.3 | 24.0 |
| mTPI2+SR | 7.8 | 12.0 | 23.5 | **30.3** | 20.1 | 6.2 | 0.0 | 4.1 | 24.5 | **57.8** | 13.4 | 0.2 | 14.9 |
| Scenario 6: High Toxicity | | | | | | | | | | | | | |
| c-CFBD | 45.4 | **39.2** | 10.3 | 3.2 | 1.5 | 0.4 | 37.6 | **49.3** | 12.9 | 0.0 | 0.0 | 0.0 | 13.1 |
| CFBD | 43.0 | **44.0** | 8.3 | 2.9 | 1.4 | 0.4 | 39.0 | **53.8** | 7.1 | 0.0 | 0.0 | 0.0 | 14.1 |
| Default BOIN | 34.5 | **35.1** | 20.4 | 7.4 | 2.2 | 0.5 | 31.3 | **46.8** | 15.9 | 1.7 | 0.2 | 0.0 | 23.6 |
| BOIN+SR | 38.3 | **27.8** | 20.5 | 9.3 | 3.3 | 0.7 | 30.2 | **40.6** | 19.3 | 5.5 | 0.9 | 0.0 | 15.4 |
| Default mTPI2 | 33.5 | **35.2** | 20.5 | 7.8 | 2.5 | 0.6 | 30.9 | **46.8** | 16.3 | 2.0 | 0.3 | 0.0 | 23.7 |
| mTPI2+SR | 37.5 | **28.0** | 21.1 | 9.2 | 3.4 | 0.8 | 29.1 | **40.9** | 21.0 | 5.2 | 0.8 | 0.0 | 15.4 |

*Acronyms: BOIN+SR, Default BOIN design with additional stopping rule 4);*
  *mTPI2+SR, Default mTPI2 design with additional stopping rule 4);*

## 2.2 Simulation study for two-agent trials

The seven scenarios for the two-agent trial are the same as those in Lee et al. (2017), as summarized in table 3. In each set of simulations, 10,000 trials were generated considering more dose combinations in two-agent trials. For the stopping rules, the minimum sample size, $n_{min}$, is 10 and the maximum, $n_{max}$, is 50. The thresholds, denoted as $r_1$ and $r_2$ in the stopping rule, are chosen to be 0.5 and 0.95, respectively. The target level ($\theta_0$) is 0.20 and the maximum tolerated toxicity ($\theta_0 + \delta_0$) is 0.25. These settings align with those of Lee et al. (2017) to facilitate comparison.

*Table 3: True toxicity rates for the seven scenarios examined, assuming four dose levels in each testing agents in each scenario. The rows under each scenario represent the dose levels of Agent A, and the columns represent the dose levels of Agent B. A target toxicity of 20% is assumed, with MTD in each scenario bolded. Note that there are no MTDs in Scenario D.*

| Scenario | Agent A | Agent B | | | |
|---|---|---|---|---|---|
| | | 1 | 2 | 3 | 4 |
| A | 1 | 0.04 | 0.10 | 0.16 | 0.22 |
| | 2 | 0.08 | 0.14 | **0.20** | 0.26 |
| | 3 | 0.12 | 0.18 | 0.24 | 0.30 |
| | 4 | 0.16 | 0.22 | 0.28 | 0.34 |
| B | 1 | 0.02 | 0.05 | 0.08 | 0.11 |
|   | 2 | 0.04 | 0.07 | 0.10 | 0.13 |
|   | 3 | 0.06 | 0.09 | 0.12 | 0.15 |
|   | 4 | 0.08 | 0.11 | 0.14 | **0.17** |
| C | 1 | 0.10 | 0.25 | 0.40 | 0.55 |
|   | 2 | **0.20** | 0.35 | 0.50 | 0.65 |
|   | 3 | 0.30 | 0.45 | 0.60 | 0.75 |
|   | 4 | 0.40 | 0.55 | 0.70 | 0.85 |
| D | 1 | 0.44 | 0.50 | 0.56 | 0.62 |
|   | 2 | 0.48 | 0.54 | 0.60 | 0.66 |
|   | 3 | 0.52 | 0.58 | 0.64 | 0.70 |
|   | 4 | 0.56 | 0.62 | 0.68 | 0.74 |
| E | 1 | 0.08 | 0.09 | 0.10 | 0.11 |
|   | 2 | 0.18 | 0.19 | **0.20** | 0.21 |
|   | 3 | 0.28 | 0.29 | 0.30 | 0.31 |
|   | 4 | 0.29 | 0.30 | 0.31 | 0.41 |
| F | 1 | 0.12 | 0.16 | 0.44 | 0.50 |
|   | 2 | 0.13 | 0.18 | 0.45 | 0.52 |
|   | 3 | 0.14 | **0.20** | 0.46 | 0.54 |
|   | 4 | 0.15 | 0.22 | 0.47 | 0.55 |
| G | 1 | 0.01 | 0.04 | 0.06 | 0.10 |
|   | 2 | 0.02 | 0.10 | 0.15 | 0.30 |
|   | 3 | 0.03 | 0.15 | 0.30 | 0.50 |
|   | 4 | 0.04 | **0.20** | 0.45 | 0.80 |

The simulation results are presented in Figures 1, 2, and 3, and the numerical values of the result are detailed in Appendix Tables A.1 and A.2. Comparing the proposed c-CFBD to the CFBD design, we observed that in general, the operating characteristics of both designs are almost the same for scenarios A and D. In scenario B, regardless of whether the calibration is applied, the dose recommendation within 1-2 percent of the target is 0 because the true toxicity rates of all doses are below the target toxicity rate. The dose recommendation within 3-5 percent and 6-10 percent of the target increases by 2.5% and 7.5%, respectively. The percentage of patients assigned to the dose within 1-10 percent of the target increased by 6.1%, indicating a lower risk of recommending a subtherapeutic dose. Moreover, after calibration, the trials require an average of 3.2 fewer patients in the c-CFBD design than CFBD. In scenario C, although there is a reduction in the average sample size and a slightly higher dose recommendation within 1-2 percent of the target, the recommendation of a dose within 1-5 or 1-10 percent of the target dropped slightly. In scenario

E, there is a more frequent dose recommendation and a higher dose allocation within 1-10 percent of the target after applying calibration. In scenario F, although the dose allocation and recommendation remain the same with or without calibration, an average of 4.7 patients are saved. Finally, in scenario G, where the benefit of calibration is most significant among all scenarios, there is a 14.3% increase in recommending the dose within 1-2 percent of the target, a 5.4 reduction in the average number of patients enrolled, and 5.6% more patients allocated to the dose within 1-2 percent of the target.

Comparing the c-CFBD design to BOIN and mTPI2/Keyboard designs without early stopping rule, the proposed method also demonstrates favorable performance in general in the combined agent setting. Specifically, the proposed c-CFBD design outperforms those designs in both percent recommendation and dose allocation in Scenarios A, E, and G, with a higher percentage of dose recommendations and patient allocations falling within 1-2 percent of the target. The accumulated values (within 1 to 5 percent, 1-10 percent, and any percentage of the target) also show favorable results for the proposed design. Furthermore, the design performs comparably to the others in percent allocation but outperformed them in dose allocation in Scenario C. It also demonstrates robustness in Scenarios D and F, only showing inferior performance in dose allocation compared to the other designs in Scenario 2. Lastly, a smaller expected sample size is observed in the proposed c-CFBD and the CFBD designs as the early stopping rule is not incorporated in the BOIN and mTPI2/Keyboard designs. After adding the early stopping rule to BOIN and mTPI-2/Keyboard designs, we observed that there is slight saving of sample sizes, i.e., an average of 5 patients across all seven scenarios (with roughly 6 patients saved in A, 1 in B, 8 in C, 1 in D, 7 in E, 7 in F, and 8 in G) are observed with the early stopping rule. However, the proposed c-CFBD again demonstrates superior performance in both percent of recommendation and patient allocation in Scenarios A, D, E, F, and G, and also shows better patient allocation in Scenario C.

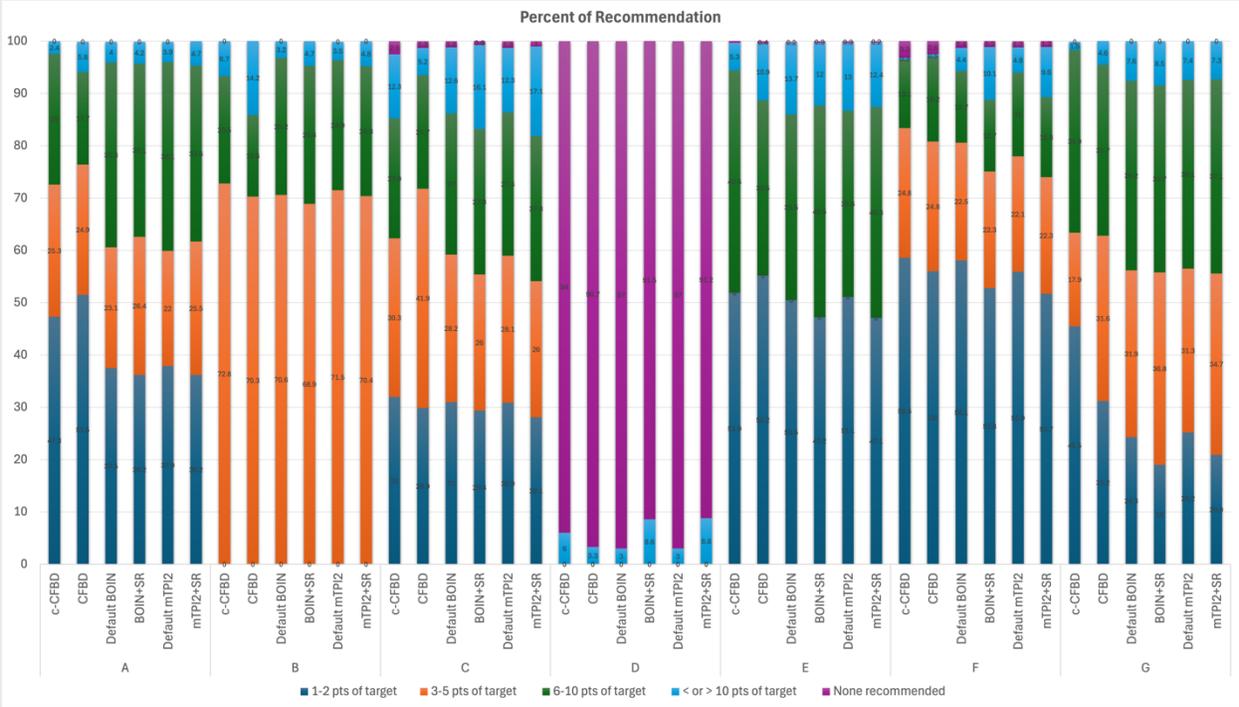

Figure 1: Stacked barplot comparing percentage of recommendation of the proposed c-CFBD, the CFBD, BOIN without and with early stopping rule 4 (default BOIN and BOIN+SR, respectively), and mTPI2/Keyboard without and with early stopping rule 4 (default mTPI2 and mTPI2+SR, respectively). The "x-y pts of target" in legend means x-y percent around the target toxicity, and the A to G on the x-axis refers to Scenarios A to G.

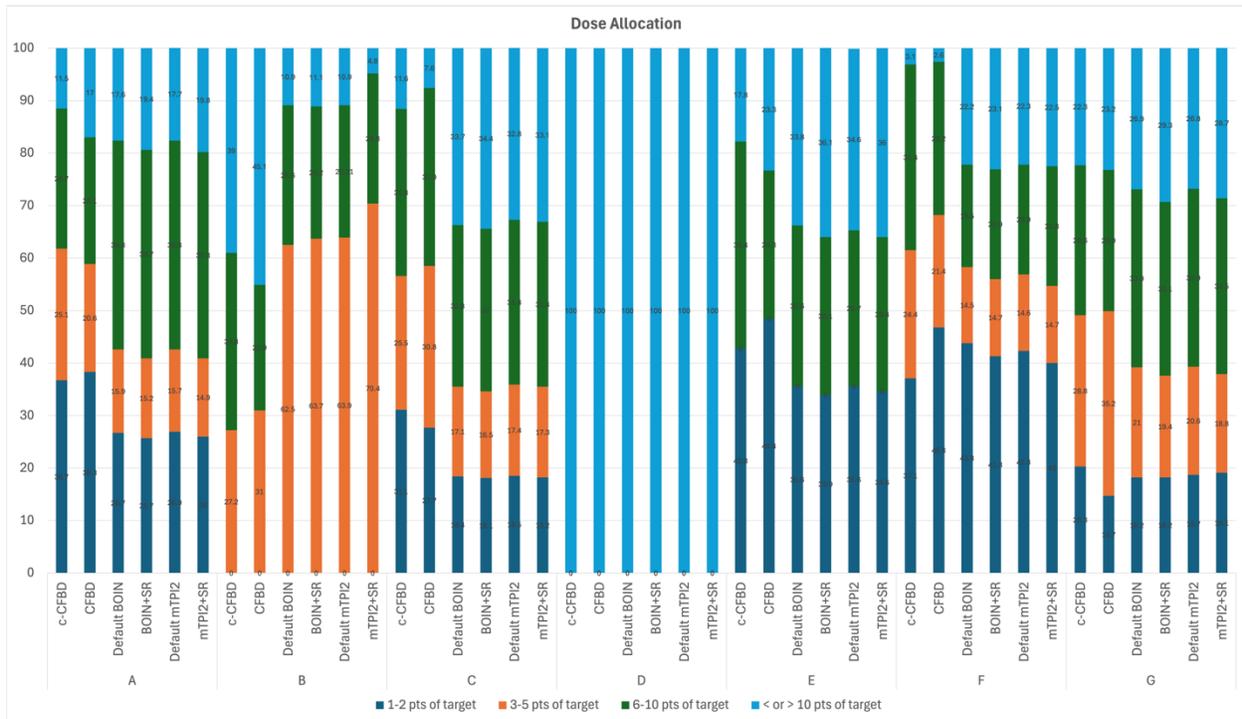

Figure 2: Stacked barplot comparing dose allocation of the proposed c-CFBD, the CFBD, BOIN without and with early stopping rule 4 (default BOIN and BOIN+SR, respectively), and mTPI2/Keyboard without and with early stopping rule 4 (default mTPI2 and mTPI2+SR, respectively). The "x-y pts of target" in legend means x-y percent around the target toxicity, and the A to G on the x-axis refers to Scenarios A to G.

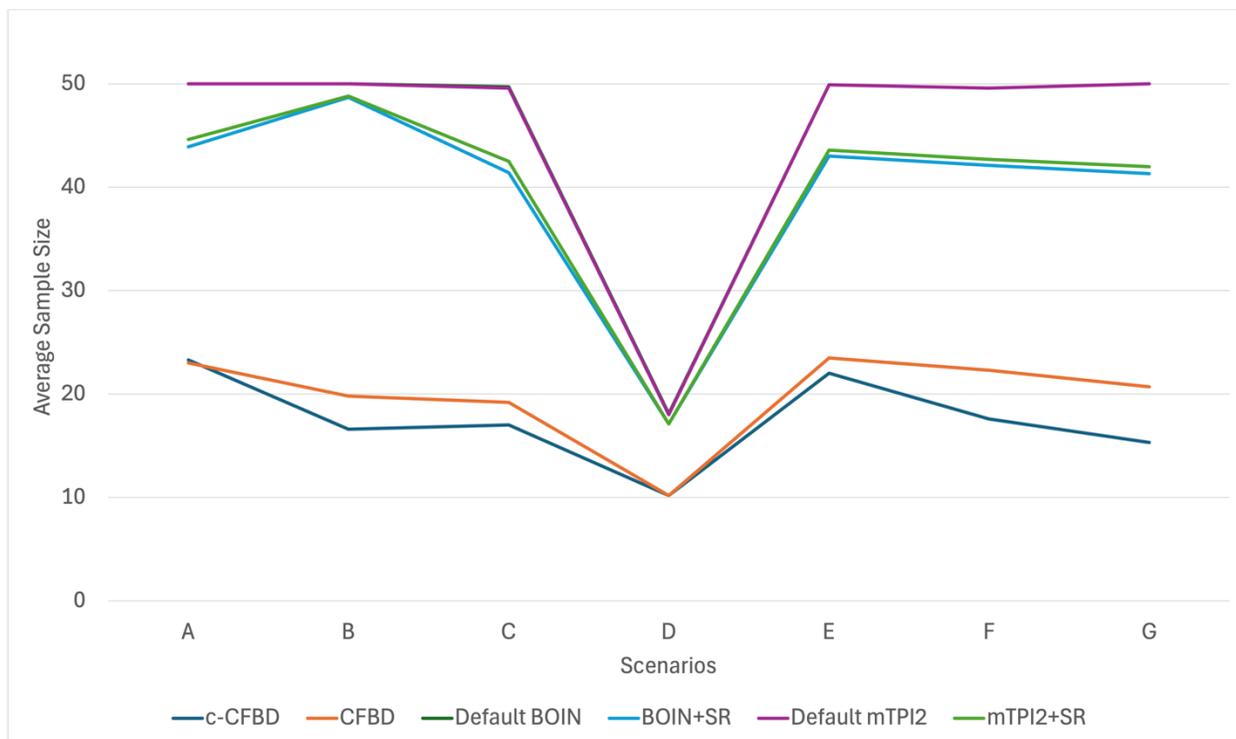

Figure 3: Plot of average sample sizes of each method under comparison across Scenarios A to G. Note that the line of Default mTPI2 (purple) overlays the line of Default BOIN (dark green).

## 3  Conclusion and discussion

In this paper, we propose an easy-to-implement method to deal with the inconsistency between effective sample size and true dose assignment to overcome the over/undershooting problem. Our goal is to identify the MTD as frequently as possible while allocating a more concentrated dose on the true MTD and minimizing the average sample size. The simulation results indicated that, except in high toxicity scenarios, calibration could enhance the performance of the Bayesian decision-theoretic design. This improvement could be in the form of more precise dose recommendations, a more concentrated allocation of doses to the MTD, or a reduction in the average sample size. Therefore, in high toxicity trials, the calibration method should be used cautiously. Moreover, the proposed calibration also demonstrated favorable performance compared to celebrated Phase I clinal trial designs such as BOIN and mTPI2/Keyboard designs. It might be more advantageous to rely on the original method if patient safety and precise dose recommendations are the highest priorities. Otherwise, based on the operating characteristic in our simulation study, c-CFBD could reduce the risk of assigning patients to subtherapeutic doses, recommending subtherapeutic doses, and help save the number of enrolled patients.

Fan et al. (2025) provided an extensive review of CFBD, along with a software package and an interactive web application developed for running simulations and determining the MTD in both single-drug and multi-drug combination trials. To facilitate the use of the proposed c-CFBD design, we integrated a new feature into our existing web application that allows users to enable or disable calibration. This enhancement allows users to simulate both the original CFBD design and the modified c-CFBD version, facilitating direct comparison of their performance metrics. By incorporating this functionality, the web application now offers researchers and practitioners a user-friendly platform to access and evaluate the improved design, available at https://cisd-stanford.shinyapps.io/CurveFreeDesigns/. This update significantly broadens the applicability of our findings.

# References

Braun, T. M., & Jia, N. (2013). A generalized continual reassessment method for two-agent phase I trials. Statistics in biopharmaceutical research, 5(2), 105-115.

Fan, S. K., Lu, Y., & Wang, Y. G. (2012). A simple Bayesian decision-theoretic design for dose-finding trials. Statistics in Medicine, 31(28), 3719-3730.

Lee, B. L., Fan, S. K., & Lu, Y. (2017). A curve-free Bayesian decision-theoretic design for two-agent phase I trials. Journal of biopharmaceutical statistics, 27(1), 34-43.

Lin, R., & Yin, G. (2017). Bayesian optimal interval design for dose finding in drug-combination trials. Statistical methods in medical research, 26(5), 2155-2167.

Liu, S., & Yuan, Y. (2015). Bayesian optimal interval designs for phase I clinical trials. Journal of the Royal Statistical Society: Series C: Applied Statistics, 507-523.

Morita, S., Thall, P. F., & Müller, P. (2008). Determining the effective sample size of a parameter prior. Biometrics, 64, 595–602.

Fan, S. K., Lee, B. L., Lu, Y., & Xu, J. (2024). Monotonic dose response assumption and curve-free designs for phase I dose finding trials. In J. Ye, D. Chen, W. Zhou, Q. Deng, & J. C. Cappelleri (Eds.), Dose finding and beyond in biopharmaceutical development. Springer. https://doi.org/10.1007/978-3-031-67110-4

Yan, F., Mandrekar, S. J., & Yuan, Y. (2017). Keyboard: a novel Bayesian toxicity probability interval design for phase I clinical trials. Clinical Cancer Research, 23(15), 3994-4003.

Yin, G., & Yuan, Y. (2009). A latent contingency table approach to dose finding for combinations of two agents. Biometrics, 65(3), 866-875.

Yin, G., & Yuan, Y. (2009). Bayesian dose finding in oncology for drug combinations by copula regression. Journal of the Royal Statistical Society Series C: Applied Statistics, 58(2), 211-224.
# Appendix

Tables A.1 and A.2 below shows the results of Figures 1, 2, and 3, the percentage of recommendation dose allocation, and average sample size, for two-agents setting, presented in Section 3.

*Table A.1: Results comparing percentage of recommendation and average sample size (column labeled by E(n)) of the proposed c-CFBD, the CFBD design of Lee et al. (2017), BOIN without and with stopping rule 4 (labeled as default BOIN and BOIN+SR, respectively), and mTPI2/Keyboard without and with stopping rule 4 (labeled as Default mTPI2 and mTPI2+SR). For columns 4 to 6, the cumulative percent of recommendation are shown in the bracket.*

|  |  | Percentage of recommendation |  |  |
|---|---|---|---|---|

| Scenario | Design | 1-2 pts of target | 3-5 pts of target (1-5 pts of target) | 6-10 pts of target (1-10 pts of target) | < or > 10 pts of target (all pts from target) | None recommended | E(n) |
|---|---|---|---|---|---|---|---|
| A | c-CFBD | 47.3 | 25.3 (72.7) | 25.0 (97.6) | 2.4 (100.0) | 0.0 | 23.3 |
|   | CFBD | 51.5 | 24.9 (76.4) | 17.7 (94.1) | 5.8 (100.0) | 0.0 | 23.0 |
|   | Default BOIN | 37.5 | 23.1 (60.7) | 35.3 (96.0) | 4.0 (100.0) | 0.0 | 50.0 |
|   | BOIN+SR | 36.2 | 26.4 (62.6) | 33.1 (95.8) | 4.2 (100.0) | 0.0 | 43.9 |
|   | Default mTPI2 | 37.9 | 22.0 (59.9) | 36.1 (96.0) | 3.9 (100.0) | 0.0 | 50.0 |
|   | mTPI2+SR | 36.2 | 25.5 (61.7) | 33.6 (95.3) | 4.7 (100.0) | 0.0 | 44.6 |
| B | c-CFBD | 0.0 | 72.8 (72.8) | 20.5 (93.3) | 6.7 (100.0) | 0.0 | 16.6 |
|   | CFBD | 0.0 | 70.3 (70.3) | 15.6 (85.8) | 14.2 (100.0) | 0.0 | 19.8 |
|   | Default BOIN | 0.0 | 70.6 (70.6) | 26.2 (96.8) | 3.2 (100.0) | 0.0 | 50.0 |
|   | BOIN+SR | 0.0 | 68.9 (68.9) | 26.4 (95.3) | 4.7 (100.0) | 0.0 | 48.7 |
|   | Default mTPI2 | 0.0 | 71.5 (71.5) | 24.9 (96.5) | 3.5 (100.0) | 0.0 | 50.0 |
|   | mTPI2+SR | 0.0 | 70.4 (70.4) | 24.8 (95.2) | 4.8 (100.0) | 0.0 | 48.8 |
| C | c-CFBD | 32.0 | 30.3 (62.3) | 22.9 (85.2) | 12.3 (97.5) | 2.5 | 17.0 |
|   | CFBD | 29.9 | 41.9 (71.8) | 21.7 (93.5) | 5.2 (98.7) | 1.3 | 19.2 |
|   | Default BOIN | 31.0 | 28.2 (59.2) | 27.0 (86.2) | 12.6 (98.8) | 1.2 | 49.7 |
|   | BOIN+SR | 29.4 | 26.0 (55.3) | 27.8 (83.1) | 16.1 (99.2) | 0.8 | 41.4 |
|   | Default mTPI2 | 30.9 | 28.1 (59.1) | 27.4 (86.5) | 12.3 (98.7) | 1.3 | 49.6 |
|   | mTPI2+SR | 28.1 | 26.0 (54.1) | 27.8 (81.9) | 17.1 (99.0) | 1.0 | 42.5 |
| D | c-CFBD | 0.0 | 0.0 (0.0) | 0.0 (0.0) | 6.0 (6.0) | 94.0 | 10.2 |
|   | CFBD | 0.0 | 0.0 (0.0) | 0.0 (0.0) | 3.3 (3.3) | 96.7 | 10.2 |
|   | Default BOIN | 0.0 | 0.0 (0.0) | 0.0 (0.0) | 3.0 (3.0) | 97.0 | 18.1 |
|   | BOIN+SR | 0.0 | 0.0 (0.0) | 0.0 (0.0) | 8.6 (8.6) | 91.5 | 17.1 |
|   | Default mTPI2 | 0.0 | 0.0 (0.0) | 0.0 (0.0) | 3.0 (3.0) | 97.0 | 18.0 |
|   | mTPI2+SR | 0.0 | 0.0 (0.0) | 0.0 (0.0) | 8.8 (8.8) | 91.2 | 17.1 |
| E | c-CFBD | 51.9 | 0.0 (51.9) | 42.5 (94.0) | 5.3 (99.3) | 0.7 | 22.0 |
|   | CFBD | 55.2 | 0.0 (55.2) | 33.5 (88.7) | 10.9 (99.6) | 0.4 | 23.5 |
|   | Default BOIN | 50.5 | 0.0 (50.5) | 35.5 (86.0) | 13.7 (99.8) | 0.2 | 49.9 |
|   | BOIN+SR | 47.2 | 0.0 (47.2) | 40.5 (87.7) | 12.0 (99.7) | 0.3 | 43.0 |
|   | Default mTPI2 | 51.1 | 0.0 (51.1) | 35.6 (86.6) | 13.0 (99.7) | 0.3 | 49.9 |
|   | mTPI2+SR | 47.1 | 0.0 (47.1) | 40.3 (87.4) | 12.4 (99.8) | 0.2 | 43.6 |
| F | c-CFBD | 58.6 | 24.8 (83.3) | 13.1 (96.4) | 0.3 (96.7) | 3.3 | 17.6 |
|   | CFBD | 56.0 | 24.8 (80.8) | 16.2 (96.9) | 0.5 (97.4) | 2.6 | 22.3 |
|   | Default BOIN | 58.1 | 22.5 (80.5) | 13.7 (94.2) | 4.4 (98.7) | 1.4 | 49.6 |
|   | BOIN+SR    | 52.8 | 22.3 (75.0) | 13.7 (88.7) | 10.1 (98.8) | 1.2 | 42.1 |
|   | Default mTPI2 | 55.9 | 22.1 (78.0) | 16.0 (94.0) | 4.8 (98.8) | 1.2 | 49.6 |
|   | mTPI2+SR   | 51.7 | 22.3 (74.0) | 15.3 (89.3) | 9.6 (98.8) | 1.2 | 42.7 |
| G | c-CFBD     | 45.5 | 17.9 (63.5) | 34.9 (98.4) | 1.6 (100.0) | 0.0 | 15.3 |
|   | CFBD       | 31.2 | 31.6 (62.7) | 32.7 (95.4) | 4.6 (100.0) | 0.0 | 20.7 |
|   | Default BOIN | 24.3 | 31.9 (56.2) | 36.2 (92.4) | 7.6 (100.0) | 0.0 | 50.0 |
|   | BOIN+SR    | 19.0 | 36.8 (55.8) | 35.7 (91.5) | 8.5 (100.0) | 0.0 | 41.3 |
|   | Default mTPI2 | 25.2 | 31.3 (56.1) | 36.1 (92.6) | 7.4 (100.0) | 0.0 | 50.0 |
|   | mTPI2+SR   | 20.9 | 34.7 (55.6) | 37.1 (92.7) | 7.3 (100.0) | 0.0 | 42.0 |

*Acronyms: BOIN+SR, Default BOIN design with additional stopping rule 4);*
*mTPI2+SR, Default mTPI2 design with additional stopping rule 4);*

*Table A.2: Results comparing dose allocation of the of the proposed c-CFBD, the CFBD design of Lee et al. (2017), BOIN without and with stopping rule 4 (labeled as default BOIN and BOIN+SR, respectively), and mTPI2/Keyboard without and with stopping rule 4 (labeled as Default mTPI2 and mTPI2+SR). For columns 4 to 6, the cumulative percent of dose allocation are shown in the bracket.*

| Scenario | Design | Dose allocation | | | |
|---|---|---|---|---|---|
|   |   | 1-2 pts of target | 3-5 pts of target (1-5 pts of target) | 6-10 pts of target (1-10 pts of target) | < or > 10 pts of target (all pts from target) |
| A | c-CFBD | 36.7 | 25.1 (61.80) | 26.7 (88.5) | 11.5 (100.0) |
|   | CFBD | 38.3 | 20.6 (58.90) | 24.1 (83.0) | 17.0 (100.0) |
|   | Default BOIN | 26.7 | 15.9 (42.58) | 39.8 (82.4) | 17.6 (100.0) |
|   | BOIN+SR | 25.7 | 15.2 (40.9) | 39.7 (80.6) | 19.4 (100.0) |
|   | Default mTPI2 | 26.9 | 15.7 (42.53) | 39.8 (82.3) | 17.7 (100.0) |
|   | mTPI2+SR | 26.0 | 14.9 (40.9) | 39.3 (80.2) | 19.8 (100.0) |
| B | c-CFBD | 0.0 | 27.2 (27.20) | 33.8 (61.0) | 39.0 (100.0) |
|   | CFBD | 0.0 | 31.0 (31.00) | 23.9 (54.9) | 45.1 (100.0) |
|   | Default BOIN | 0.0 | 62.5 (62.52) | 26.6 (89.1) | 10.9 (100.0) |
|   | BOIN+SR | 0.0 | 63.7 (63.7) | 25.2 (88.9) | 11.1 (100.0) |
|   | Default mTPI2 | 0.0 | 63.9 (63.91) | 25.21 (89.1) | 10.9 (100.0) |
|   | mTPI2+SR | 0.0 | 70.4 (70.4) | 24.8 (95.2) | 4.8 (100.0) |
| C | c-CFBD | 31.1 | 25.5 (56.6) | 31.8 (88.4) | 11.6 (100.0) |
|   | CFBD | 27.7 | 30.8 (58.5) | 33.9 (92.4) | 7.6 (100.0) |
|   | Default BOIN | 18.4 | 17.1 (35.5) | 30.8 (66.3) | 33.7 (100.0) |
|   | BOIN+SR  | 18.1 | 16.5 (34.6) | 31.0 (65.7) | 34.4 (100.0) |
|   | Default  | 18.5 | 17.4 (35.7) | 31.4 (67.2) | 32.8 (100.0) |
|   | mTPI2+SR | 18.2 | 17.3 (35.5) | 31.4 (66.9) | 33.1 (100.0) |
| D | c-CFBD   | 0.0  | 0.0 (0.0)   | 0.0 (0.0)   | 100.0 (100.0) |
|   | CFBD     | 0.0  | 0.0 (0.0)   | 0.0 (0.0)   | 100.0 (100.0) |
|   | Default BOIN | 0.0 | 0.0 (0.0) | 0.0 (0.0) | 100.0 (100.0) |
|   | BOIN+SR  | 0.0  | 0.0 (0.0)   | 0.0 (0.0)   | 8.6 (8.6)    |
|   | Default  | 0.0  | 0.0 (0.0)   | 0.0 (0.0)   | 100.0 (100.0) |
|   | mTPI2+SR | 0.0  | 0.0 (0.0)   | 0.0 (0.0)   | 100.0 (100.0) |
| E | c-CFBD   | 42.8 | 0.0 (42.8)  | 39.4 (82.2) | 17.8 (100.0) |
|   | CFBD     | 48.4 | 0.0 (48.4)  | 28.3 (76.7) | 23.3 (100.0) |
|   | Default BOIN | 35.6 | 0.0 (35.6) | 30.6 (66.2) | 33.8 (100.0) |
|   | BOIN+SR  | 33.9 | 0.0 (33.9)  | 30.1 (63.9) | 36.1 (100.0) |
|   | Default  | 35.6 | 0.0 (35.6)  | 29.7 (65.4) | 34.6 (100.0) |
|   | mTPI2+SR | 34.6 | 0.0 (34.6)  | 29.4 (64.0) | 36.0 (100.0) |
| F | c-CFBD   | 37.1 | 24.4 (61.5) | 35.4 (96.9) | 3.10 (100.0) |
|   | CFBD     | 46.8 | 21.4 (68.2) | 29.2 (97.4) | 2.60 (100.0) |
|   | Default BOIN | 43.8 | 14.5 (58.3) | 19.5 (77.8) | 22.2 (100.0) |
|   | BOIN+SR  | 41.3 | 14.7 (56.0) | 20.9 (76.9) | 23.1 (100.0) |
|   | Default  | 42.3 | 14.6 (56.8) | 20.9 (77.7) | 22.3 (100.0) |
|   | mTPI2+SR | 40.0 | 14.7 (54.7) | 22.8 (77.5) | 22.5 (100.0) |
| G | c-CFBD   | 20.3 | 28.8 (49.1) | 28.6 (77.7) | 22.3 (100.0) |
|   | CFBD     | 14.7 | 35.2 (49.9) | 26.9 (76.8) | 23.2 (100.0) |
|   | Default BOIN | 18.2 | 21.0 (39.2) | 33.9 (73.1) | 26.9 (100.0) |
|   | BOIN+SR  | 18.2 | 19.4 (37.6) | 33.1 (70.7) | 29.3 (100.0) |
|   | Default  | 18.7 | 20.6 (39.3) | 33.9 (73.2) | 26.8 (100.0) |
|   | mTPI2+SR | 19.1 | 18.8 (37.8) | 33.5 (71.3) | 28.7 (100.0) |

*Acronyms: BOIN+SR, Default BOIN design with additional stopping rule 4);*
    *mTPI2+SR, Default mTPI2 design with additional stopping rule 4);*